# Design of Maximum-Gain Dielectric Lens Antenna via Phase Center Analysis

Md Khadimul Islam
*Department of Electrical & Computer Engineering*
*Florida International University*
Miami, FL, USA
misla081@fiu.edu

Arjuna Madanayake
*Department of Electrical & Computer Engineering*
*Florida International University*
Miami, FL, USA
amadanay@fiu.edu

Shubhendu Bhardwaj
*Department of Electrical & Computer Engineering*
*Florida International University*
Miami, FL, USA
sbhardwa@fiu.edu

*Abstract*—In this work, a method is presented to maximize the obtained gain from millimeter-wave (mm-wave) lens antennas using phase center analysis. Commonly, for designing a lens antenna, the lens is positioned just on top of the antenna element which is not capable of providing the maximum gain/aperture efficiency. A novel solution method is proposed where the lens will be placed at a distance calculated using phase center analysis to produce the maximum gain from the system. A mm-wave microstrip antenna array is designed and the proposed method is applied for the gain enhancement. Simulation results suggest that the propose scheme obtains around 25% gain enhancement compared to the traditional method.

## I. INTRODUCTION

Communications at mm-wave bands have been proven as a promising solution for the bandwidth demand for the next generation wireless communication system. In particular, wide range of mm-wave band (28 GHz -300 GHz) enables high-bandwidth communication via abundance of spectrum. This band is being used in numerous applications, for instance, radar [1], satellite [2], imaging system [3] etc. In such applications, high power beam generation is required which involves antenna systems capable of generating highly directive beams. The antenna systems serving this purpose suffer from losses due to the surface modes. However, making use of lens on top the antenna provides a solution to that problem. Beside reduction of loss due to the surface mode, lens adds the thermal stability and mechanical rigidity to the system [4].

Path loss is another key issue in mm-wave communication system [5]. When a highly directive beam encounters an obstacle in its way of propagation, the signal undergoes a huge loss of power intensity. Multiple simultaneous beams allows a partial solution to the problem of beam blockage. Lens antennas with focal plane-array beamforming furnishes an efficient scheme for obtain multiple simultaneous beams. Sharp beams are directed across several angles so that path loss and occlusions both can be avoided towards establishing a wireless link through the mm-wave channel. Multiple generated beams using the phase synchronization can provide multiple transmission paths to avoid the obstacles and results in reduction of path loss [6]–[8]. However, the generation of hybrid beams requires phase shifter in the analog processing stage which makes the beam forming circuitry complex [9]. Lens can be used here as phase shifter to make system simple while keeping the gain very high.

Studies in this field show that lenses are generally placed on antenna top surface without any gap in between, to reduce the loss, minimize the complexity or to produce highly directive sharp beams. Nevertheless, such arrangement of lens positioning on the top surface of the antenna is not capable of providing the optimum output compared to the theoretically calculated gain value [8]. It is assumed that the non-spherical wavefront from the antenna at the focal point of the lens is hindering the system to provide the optimum gain. To attain the maximum gain, the placement of the lens and the antenna should be adjusted. Hence the motivation of this work is the maximization of the gain produced by a system by proper placement of the antenna and lens.

To this end, we proposed the use of phase center analysis to improve the gain value from a system. Generally, phase center can be simply defined by denoting a point where the phase front is well formed [10]. Here, we use a separation between the top surface of the antenna and the bottom surface of the lens to allow the signal to create a well formed phase front. This separation is determined from the phase center calculation. The application of our proposed method is tested for the microstrip patch antenna array. A satisfactory gain enhancement is obtained from the system.

## II. PHASE CENTER

According to the IEEE standard (2.270), the phase center can be defined as, the location of a point associated with an antenna such that, if it is taken as the center of a sphere whose radius extends into the far-field, the phase of a given field component over the surface of the radiation sphere is essentially constant, at least over that portion of the surface where the radiation is significant [11]. Ideally, the phase should be constant over the measurement surface on the radiation sphere which in practice is not always true. So a phase function needs to be defined to account for the phase variation on the considered surface. The minimization of this

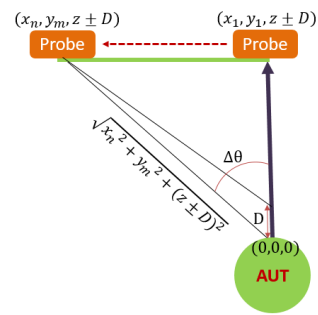

Fig. 1: Determination of the phase center of an antenna under test (AUT).

function will provide the location of the phase center. Least Square Fit method can be used to minimize the phase function over the surface [12]. Assuming the antenna is at origin as shown in Fig. 1, the probe position will be varied on the xy

$$S(D) = \sum_n \sum_m ((\varphi_{ideal}(x_n, y_m, z \pm D) - \varphi_{measured}(x_n, y_m, \pm D))^2 \quad (1)$$

plane for different value of height from the origin, $D$. Then the function which needs to be minimized will be,

where, $S$ is phase function, $\varphi_{ideal}$ is the reference phase value and $\varphi_{measured}$ is the phase value at different points on the xy plane. At the minimum value of $S$, the phase front of the signal will be well formed and the phase error will be in the limit of minimum phase error (22.5°) [13]. The area of the measurement plane is determined by a solid angle, $\Delta\theta$. The solid angle ($\Delta\theta$) needs to be taken in such a way so that most of the radiated power passes through it. The position of the phase center will be at the height ($D$) where the phase function

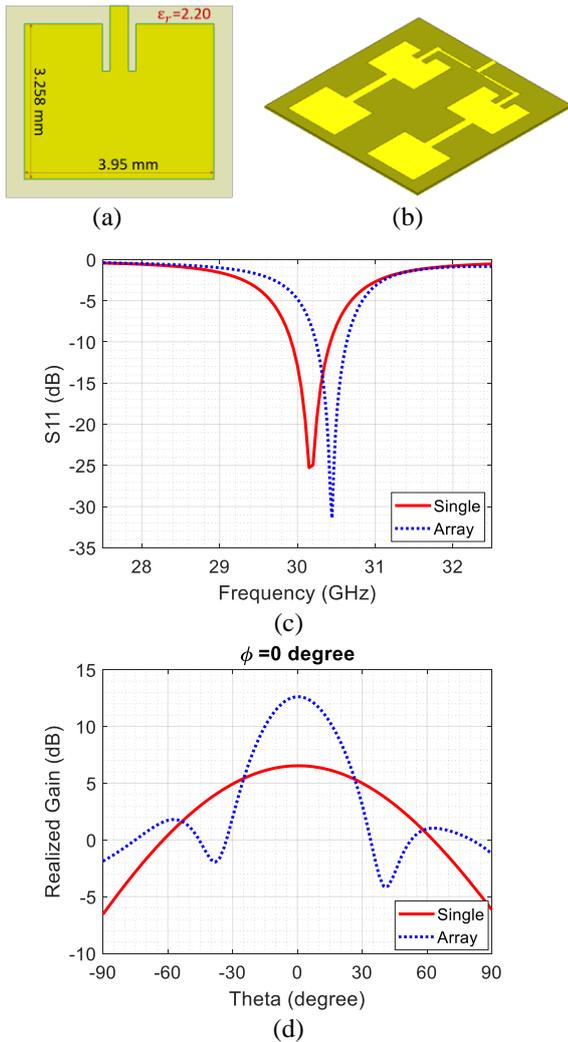

Fig. 2: Design of the single patch (a) and array antenna (b). S-parameters (c) and realized gain (d) for both of the antennas.

($S$) is minimum.

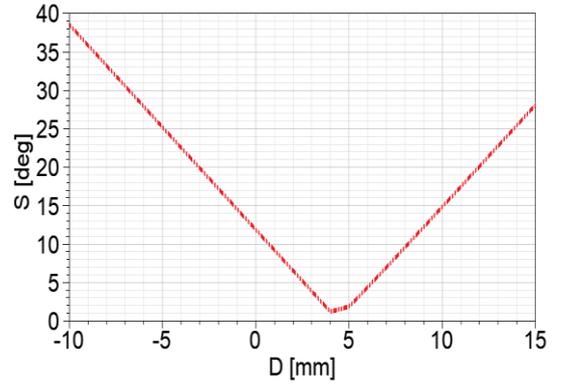

Fig. 3: Determination of phase center of the array.

### III. ANTENNA DESIGN

To validate the feasibility of our proposed method, we show the specific design of an antenna array. We designed a 2x2 microstrip patch antenna array to operate at 30.2 GHz [13]. The length and the width of the radiating elements are considered to be 3.258mm and 3.95 mm, respectively. Commercially available RT/Duroid 5880 material with the height 0.127 mm is used as the substrate. Inset feeding technique is used to feed the patch element while the corporate feeding is used to excite the array. The single patch and antenna array shown in Fig. 1 (a) and (b). Commercially available FEM solver Ansys HFSS is used to simulate the structure. A response below -10 dB at the operating frequency is observed from the s-parameter plot (Fig. 1 (c)) for both of the cases. The maximum realized gains are found to be 6.53 dB for the single antenna and 12.62 dB for the array at the resonance frequencies. The gain increment from single element to the array looks satisfactory. In the next section, we show the reader the proposed design procedure for obtaining the desired matching via phase center analysis.

### IV. PHASE CENTER CALCULATION OF THE ANTENNAS

Following the procedure described in section II, the phase centers of the antennas are calculated. For this study, 22.5° is taken as the value of the solid angle ($2\Delta\theta = 45°$) to create the measurement plane and a set of simulations are run with a range $D = -30\ mm$ to $D = 30\ mm$, to find the position of the phase center. Fig. 3 shows the plot of the phase function ($S$) vs height ($D$). From the figure, one can see that the phase function on the observation plane has minimum value at $D = 4.4\ mm$ i.e. the phase center of the array is situated at the height of 4.4 mm from the top surface of the antenna. This value of D will be used as the distance between the lens and the antenna in hope to get the maximum gain

### V. DESIGN OF THE LENS

The literature provides the evidence of the use of both of the homogeneous and inhomogeneous lenses [14]. In this study, we used the homogeneous lens. These lenses can be of various shapes, for instance, Spherical, extended hemispherical, elliptical [4], [14] etc. In the extended hemispherical lenses, the extension in the length provides the control over radiation properties, that makes it the most used type of lens. In our study, we used this extended hemispherical lens. For a hemisphere of radius $R$, the height of the extension, L can be calculated with following equation,

$$L = b\frac{1+\frac{1}{n}}{\sqrt{1-\frac{1}{n^2}}} - R \qquad (2)$$

The variable b can be calculated with the following equation,

$$b = R(1 + \frac{1}{3n^2})$$

where, $n$ is refractive index of the lens material and it can be calculated from the dielectric constant of that material ($n = \sqrt{\varepsilon_r}$).

For this work, we choose the dielectric constant to be 2.4, radius of the hemisphere to be 17.27 mm and extension height is 24.15 mm which should produce the gain of 20.3 dB. Then placed the lens on top of the antenna for further analysis (Fig. 4 (a)).

## VI. PERFORMANCE ANALYSIS OF THE LENS ANTENNA

After designing the array and the lens, we placed the lens on top of the array and analyzed the performance using HFSS. At first we placed it in such a way so that the bottom of the lens rest on the top surface of the patch array ($D = 0\ mm$). The maximum realized gain at this distance ($D = 0mm$) at the resonance frequency of 30.2GHz is seen to be 14.7 dB. A little improvement of the gain is observed but this is much lower than the desired gain value. Hence, we employ the phase center analysis for the maximization of the gain. We run another simulation using $D = 4.4\ mm$, the calculated distance of phase center. The realized gain for this case is found to be 19.4 dB at resonance frequency which provides about 25% of improvement with respect to its initial position. The observed value is also close to the theoretically calculated gain value. The realized gain includes the total antenna efficiency/losses inside the materials used in the system, hence little discrepancy between the theoretical gain and realized is expected. Fig. 4 (b) shows the gain comparison of the realized gain for the three cases, only array (no lens), array+lens at $D = 0\ mm$ and array+lens at $D = 4.4\ mm$. Further, we run some simulations using different values for $D$. Fig. 5 shows the gain vs height ($D$) plot. One can see that when the lens is not at phase center, let say a point $D = 2.42mm$, the gain is 17.1dB, which is lower than the gain value at phase center ($D = 4.4\ mm$). It is observed that, starting from the top surface of the antenna ($D = 0$), the gain is increasing and approaching the theoretical limit as the separation ($D$) reaches the phase center and then again going downwards for higher values of $D$. This is because, at phase center, the waves are acting as a point source at the focal point of the lens which results in the maximization of gain.

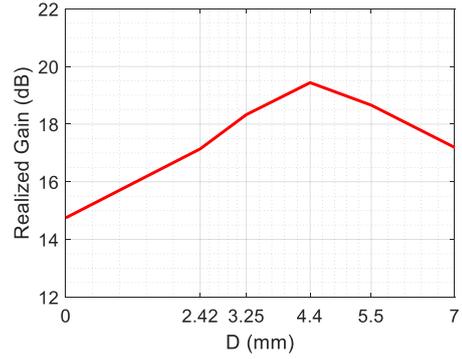

Fig. 5: Maximum realized gain at different separation, D.

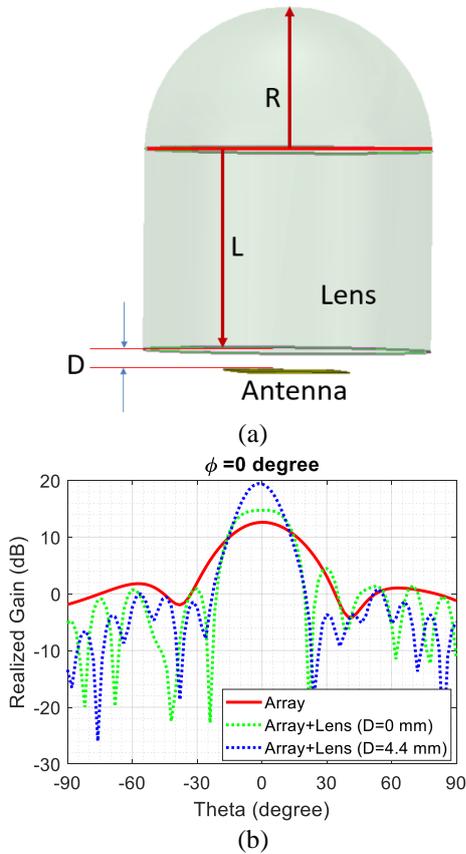

Fig. 4: Arrangement of the lens antenna (a) and comparison of the realized gain from the array with and without the lens placed on the top surface of the antenna.

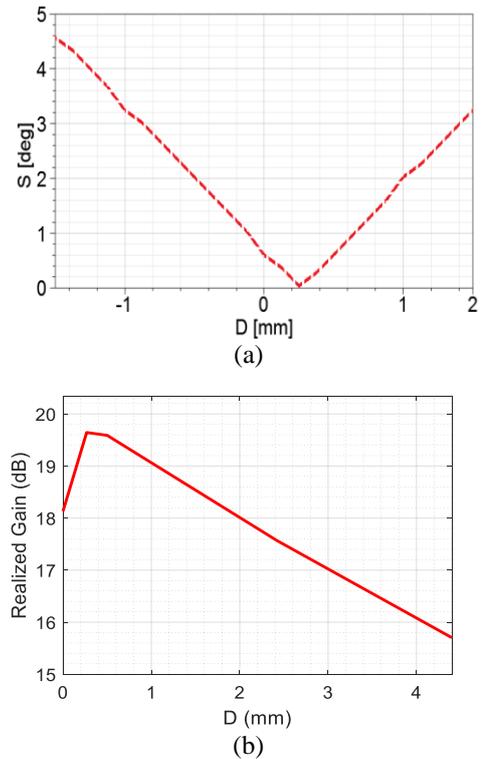

Fig. 6: (a) Determination of phase center position of the single element antenna and (b) maximum realized gain vs separation, D.

The same analysis can be conducted for the single element antenna. But in the case of single patch, the phase center resides very close to the top surface. In such case, a practically useful gain value can be obtained even if the antenna is placed touching the top surface of the antenna. To show the variation, we performed the same phase center analysis for the single element. Fig. 6 (a) shows the plot of the phase function (S) vs the separation (D). Here, the phase center is 0.25 mm which is very close to the top surface. For this case, when the lens is placed touching the top surface of the antenna, the boresight gain value is more than 18 dB which is close to the desired value. Fig. 6 (b) shows the realized gain vs separation (D) plot for single element antenna. The increment in the separation towards the calculated phase center value results in improvement of the gain. The gain reached to the maxima of gain value 19.6 dB at $D = 0.25mm$. Further increase of the separation leads to diminishing returns.

## VII. Conclusion

We proposed and showed the application of phase center analysis for optimizing the gain of lens antennas. We showed that the lens antenna will produce optimum gain while the bottom the lens is placed at well-formed phase front. The distance of that phase front is determined by the phase center analysis which uses the least square fit method to minimize the phase function. The analysis shows that an improvement of 25% in the value is obtained for a 2x2 array antenna. Though our proposed method is comparatively less effective for the single element antennas but has significant effect on the mm-wave lens antennas with array in the system.